\documentclass[a4paper, 12pt]{article}
\usepackage[top=2.54cm, bottom=2.54cm, left=2.54cm, right=2.54cm]{geometry}
\usepackage{floatrow}
\newfloatcommand{capbtabbox}{table}[][\FBwidth]

\reversemarginpar
\usepackage{todonotes}
\usepackage{xcolor}\definecolor{tum_blue}{RGB}{0, 115, 207}\colorlet{col_section }{tum_blue}

\newcommand\independent{\protect\mathpalette{\protect\independenT}{\perp}}
\def\independenT#1#2{\mathrel{\rlap{$#1#2$}\mkern2mu{#1#2}}}
\usepackage[latin1]{inputenc}
\usepackage{tikz}
\tikzset{
  treenode/.style = {shape=rectangle, rounded corners,
                     draw, align=center,
                     top color=white, bottom color=blue!20},
  root/.style     = {treenode, font=\normalsize, bottom color=blue!30},
  decision/.style      = {treenode, font=\normalsize, bottom color=red!30},
  env/.style      = {treenode, font=\normalsize},
  dummy/.style    = {circle,draw}
}
\usetikzlibrary{shapes,arrows}
\usepackage{
	booktabs
	,authblk
	,lmodern
	,multirow
	,subfigure
	,graphicx
	,amssymb
	,amsfonts
	,amsmath
	,amsthm
	,amssymb
	,stmaryrd
	,color
	,array
	,enumerate
    ,microtype
    ,geometry
    ,graphics
	,threeparttable
	,longtable
	,rotating
	,lscape
	,tabularx
	,epsfig
    ,epstopdf
	,setspace
	,titling
}

\usepackage{chngcntr}
\counterwithin{equation}{section}
\usepackage[tableposition=top]{caption}
\newcolumntype{Y}{>{\centering\arraybackslash}X}

\usepackage[multiple]{footmisc}

\usepackage{hyperref}
\usepackage{breakurl}
\hypersetup{
	pdfstartview = FitH,
	pdfauthor = {...},
	pdftitle = {...},
	pdfkeywords = {...; ...; ...; ...},
	linktocpage=true,
	colorlinks=true,      
    linkcolor=blue,       
    citecolor=blue,       
    filecolor=blue,       
    urlcolor=blue         
}

\usepackage{xurl}
\usepackage{chapterbib}
\usepackage{natbib}

\DeclareMathOperator*{\argmax}{arg\,max}
\DeclareMathOperator*{\argmin}{arg\,min}
\newtheorem{proposition}{Proposition}
\newtheorem{assumption}{Assumption}

\title{Using Probabilistic Stated Preference Analyses\\ to Understand Actual Choices \thanks{Correspondence address: Manor Road Building, Manor Road, Oxford, OX1 3UQ. email: romuald.meango@economics.ox.ac.uk. The author thanks Eleanor Dickens for invaluable research assitance. The author is grateful to Johannes Abeler, Abiga\"{i}l Adams, Manuel Arellano, Kirill Evdokimov, Max Kasy, Fran\c{c}ois Poinas, Lance Lochner, Roland Rathelot, Martin Weidner, Basit Zafar, and participants to the 7th Workshop on Subjective expectations, and the workshop on Microecoometrics and Policy evaluation of the BSE Summer Forum for stimulating discussions and useful comments at different stages of this paper. All remaining errors are mine.}}
\author{Romuald M\'{e}ango}
\affil{University of Oxford and CESifo}
\date{This version: \today}
\begin{document}

\maketitle

\begin{abstract}
    Can stated preferences help in counterfactual analyses of actual choice? This research proposes a novel approach to researchers who have access to both stated choices in hypothetical scenarios and actual choices. The key idea is to use probabilistic stated choices to identify the distribution of individual unobserved heterogeneity, even in the presence of measurement error. If this unobserved heterogeneity is the source of endogeneity, the researcher can correct for its influence in a demand function estimation using actual choices, and recover causal effects. Estimation is possible with an off-the-shelf Group Fixed Effects estimator.
\end{abstract}

\noindent \textbf{Keywords:} Stated preference; revealed preference; unobserved heterogeneity.\\

\noindent \textbf{JEL codes:} C30, C33, D84.
\clearpage

\section{Introduction}
Stated preference analyses typically present agents with discrete alternatives in several hypothetical scenarios and elicit their intended choice in each. Recent applications are wide-spanning and include publications in leading journals in economics: among other topics, education and degree choices \citep{delavande2019, arcidiacono2020}, mobility decisions \citep{gong2022, kocsar2022}, health and long-term care investments \citep{kesternich2013, ameriks2020b,boyer2020}, parental investments \citep{almaas2023}, voting behaviour \citep{delavande2015, debresser2019},  marriage preferences \citep{adams2019}, occupational choices \citep{wiswall2015, maestas2018, wiswall2018}, retirement decisions \citep{ameriks2020a} and irregular migration \citep{bah2018,meango2022}. See also a recent literature review by \cite{kocsar2023}.\footnote{A larger literature on stated choice analysis is referenced in \cite{kocsar2022}, footnotes 6 and 7. \cite{almaas2023} provides an historical perspective on the debate over the suitability of stated preference and revealed preference analyses.}  \cite{almaas2023}, taking inspiration from Orazio Attanasio's presidential address to the Econometric society, provides compelling arguments for empirical strategies integrating stated and revealed preferences. They note: `recent developments indicate that the [economics] profession is moving towards using choice data and directly observable variables in combination with \textit{stated preferences} and answers to hypothetical questions.'

The increasing interest in the stated preference approach stems from its numerous advantages over the revealed preference approach (i.e. an approach using actual choices only): it avoids assumptions about agents' belief formation and equilibrium allocation mechanisms, and is a more natural approach to questions pertaining to ex ante perceptions. Using \textit{probabilistic} stated choices, that is, the chance of choosing an alternative on a scale from 0 to 100, rather than a binary answer, provides the means for respondents to express uncertainty about their future choice and for researchers to learn about this uncertainty \citep{blass2010, meango2023}. One key advantage of stated preference analyses is that the analyst can exploit hypothetical choice scenarios to tackle the pervasive problem of endogeneity arising from omitted variable bias and identify causal effects of choice attributes.\footnote{\cite{ameriks2020b} use hypothetical choices elicited in strategic survey question to estimate some preference parameters in a fully specified structural model. The models addressed in this paper are nonparametric by nature.}  Even if not all relevant choice characteristics are observed, exogenous variations on observed characteristics induced in a survey experiment are sufficient for causal inference.

Given the increasing importance of stated preference analyses, it is essential to address an ever-recurring question: Can stated choices serve to understand actual behaviour or causal effects in actual choice environments? There is growing evidence that a carefully designed stated preference elicitation often has predictive power for actual choices in several contexts \citep[see, e.g. ][]{hurd2009, bah2018, wiswall2018, debresser2019, arcidiacono2020} but concerns remain regarding systematic biases of this method. See, for example, \cite{murphy2005} and more recently \cite{haghani2021a,haghani2021b}. The conventional approach to the above question, when one has access to both hypothetical and actual choices, remains unsophisticated. It involves assessing the correlation between stated and actual outcomes, or measuring the size of the biases between stated and average choice in sub-populations defined by observed characteristics. Some studies have combined elicited and actual choices in structural model estimation, finding some gains in estimation precision.\footnote{See, for example, \cite{vanderklaauw2012, kesternich2013, zafar2013} and an earlier literature in transport and environmental economics, surveyed in \cite{whitehead2008}.} However, counterfactual analyses using subjective choices have largely remained separate from causal inference on objective choices, since there is no consensus on remedies to biases in stated preference analyses. Two notable exceptions are \cite{briggs2020} and \cite{bernheim2022} that are discussed below.

This research proposes a novel and simple approach to researchers who have access to both stated choices in hypothetical scenarios and actual choices in a binary choice environment, and wish to estimate counterfactual parameters. The key idea is that \textit{probabilistic} stated choices not only allow for learning about respondents' uncertainty, but also provide the means to learn about the distribution of \textit{individual unobserved heterogeneity}. 
To motivate with a simple example, suppose that an analyst has access to data on intended migration choices in different hypothetical scenarios, and subsequent migration behaviour. A respondent who reports a high probability of migration in most scenarios is arguably different from a respondent who states a low probability of migration in most scenarios. Thus, the stated preferences contain some information about individual `types', and can be harnessed to discriminate among types in the population. Once these types have been identified, they may serve as control variables in a demand function estimation using individual's actual choices.

This paper provides a general framework where the distribution of multi-dimensional unobserved heterogeneity can be identified from probabilistic stated choices in different hypothetical scenarios. The approach allows for respondents to make mistakes and/or be biased about the future, as long as (1) the observable and unobservable attributes that matter in the actual choice also influence the stated choice, and, if the analyst is interested in causal inference on the demand function, (2) the unobserved heterogeneity in the stated choice contains a sub-vector that is the only source of endogeneity. Hence, stated preferences can serve as control functions. These requirements depart from the previous literature that combines both stated preference and revealed preference data while assuming that expectations about future behavior accurately portray optimal future behaviour conditional on current information \citep[see, for example,][]{vanderklaauw2012}, or from the conditions in \cite{bernheim2022} that the difference between stated and revealed demand function is stable across treatment groups.

Section \ref{sec:model} presents the econometric framework that justifies matching individuals on their stated preferences before performing a regression of actual choices on choice characteristics. It shows identification for the mean squared bias that gives an accurate assessment of the bias between stated and actual choices, counterfactual distribution functions in the spirit of \cite{chernozhukov2013}, and the average structural function \citep{blundell2004}, that serves for causal inference of the effect of choice attributes on actual choice. The case of an uni-dimensional heterogeneity serves as a motivating example. In this case, a probabilistic stated choice in only one hypothetical scenario is required. When the heterogeneity is multidimensional, say $d>1$, at least $d$ scenarios are required for identifying the unobserved heterogeneity.

Section \ref{sec:measurement_error} incorporates (classical) measurement error. It shows that even with a small number of hypothetical scenarios, the objects of interest are identified. The result rests on \cite{kotlarski1967}'s Lemma, as the identification results for nonseparable panel models in \cite{evdokimov2010} and \cite{meango2023}, respectively.

Section \ref{sec:estimation} considers estimation. It adapts the Two-Step Group Fixed Effects estimator (TSGFE) proposed by \cite{bonhomme2022}.  TSGFE exploits auxiliary moments to classify individuals into a finite number of types, based on a continuous, low-dimensional unobserved heterogeneity. The recovered types are then used in the estimation of the parameters of interest. In this context, TSGFE is a natural approach to use, as probabilistic stated choices generate identifying moments for individual unobserved heterogeneity and serve to classify individuals. The identified types correct for the influence of the unobserved heterogeneity on the parameters of interest.

Section \ref{sec:discussion} discusses the importance of the result for practitioners, and its link to the proposals from \cite{briggs2020} and \cite{bernheim2022}. Section \ref{sec:conclusion} concludes. Appendix \ref{app:testing} considers the question of whether a given survey experiment exhausts the information about the unobserved heterogeneity and derives testable implications for restrictions on the dimension of unobserved heterogeneity.

\section{Identification in the Absence of Measurement Error}\label{sec:model}
\subsection{Econometric Framework}\label{subsec:framework}
Consider an economic agent $i$, a binary choice alternative $0$ or $1$, and two consecutive periods: a time of preference elicitation and a time of decision. At the time of decision, $i$ chooses between option $0$ and option $1$ based on a threshold-crossing rule:
\begin{equation}\label{eq:choice_equation}
\begin{array}{rll}
    D_i(x) &=&I \bigl\{{S^r \left({x,\eta_i}\right) \ge \nu_i^r}\Bigr\}  \text{ and  }\\
    D_i &=& D_i(X_i)
\end{array}
\end{equation}
The notation borrows from the potential outcome framework, as $D_i(x)$ represents $i$'s choice, when the choice attributes are characterised by a vector $x$, and $D_i$ is the actual choice, as observed by the analyst at the time of decision. The potential outcomes notation emphasises the analyst's interest in `treatment effects' and counterfactual policies. The choice is binary and can be, for example, between migrating or not, following a STEM degree or not, retiring early or late. The vector-valued variable $X_i$ represents choice characteristics that are observed by the analyst at the time of decision and can be manipulated in a choice experiment. In the example of a migration decision, it could include the average wage at origin and destination and the pecuniary cost of moving. In the example of major choice, it could include average wages in STEM and non-STEM occupations, average female representation, or average success rates. 

The variable $\eta_i$ subsumes unobserved preferences over choice attributes that influence $i$'s decision, for example, $i$'s taste for migration or innate ability in math-related subjects. The dimension of $\eta$ is finite and the reader should think of $\eta$ as having a low dimension. The variable $\nu_i^r$ captures the \textit{resolved uncertainty}, a realisation of what is known as the \textit{resolvable uncertainty}. It summarises choice characteristics that are unknown to $i$ at the time of elicitation, but are revealed at the time of decision \citep[see an extended discussion of the resolvable uncertainty in ][]{blass2010}. The implicit assumption of Equation (\ref{eq:choice_equation}) that $\nu_i^r$ is separable and univariate is standard in the stated choice literature and most discrete choice models. It implies that the unobserved (dis-)utility caused by preference shocks does not vary with the observed choice characteristics and the unobserved heterogeneity. This is equivalent to the assumption of a univariate, non-separable resolvable uncertainty such that the mapping $\nu \mapsto S^r(x,n,\nu)$ is strictly increasing for each $(x,n) \in \mathcal{X} \times \mathcal{H}$, the joint support of $X_{i}$ and $\eta_i$ \citep{vytlacil2002}. The characteristics $\eta_i$ is thought of as being \textit{stable}, that is, it does not change between the time of elicitation and the time of decision, but may be correlated with the resolved uncertainty. Finally, $S^r \left({X_i,\eta_i}\right)-\nu_i^r$ can be interpreted as $i$'s perceived returns on choosing option 1 over option 0. 

Prior to their decision and at the time of elicitation, $i$ is asked to state their preference over the binary choice alternative $0$ or $1$ in a hypothetical scenario $t$ where $X_{it} = x_{it}$ and reports:
    \begin{equation}
    P_{it} = \int_\mathcal{V} I \bigl\{{S \left({x_{it},\eta_i}\right) \ge v}\Bigr\} dF_{\nu \vert X_i, \eta_i}(v \vert x_{it},\eta_i)
\end{equation}
where $F_{\nu \vert X_i, \eta_i}(v \vert x_{it},\eta_i)$ is the \textit{perceived distribution} of resolvable uncertainty at the time of elicitation. $S \left({x_{it},\eta_i,\nu_i}\right) $ is $i$'s average perceived returns to choosing option 1 over option 0 at the time of elicitation, when the choice characterisitics are $x_{it}$. As above, $\nu_i$ is assumed to be univariate and additively separable.

Beyond the fact that they have different supports, there are two main differences between probabilistic stated choices and the actual choice. First, $F_{\nu \vert X_i, \eta_i}$ is not necessarily the same as the distribution of $\nu^r$, that is, individuals can be mistaken about the distribution of future realisations of the resolvable uncertainty. Second, $S(\cdot)$ and $S^r(\cdot)$ are not necessarily the same, that is, at the time of elicitation, individuals can be mistaken about their perception of returns at the time of decision, for example if they are subject to present bias or hyperbolic discounting.

The framework will not allow distinguishing between the two sources of bias without further assumptions. Furthermore, because of the binary nature of the choice, it will not be possible to make progress on the identification of $S^r(.)$ without a normalisation of the distribution on $\nu_i^r$. Therefore, without loss of generality, one can write:
\begin{equation}\label{eq:definition_D}
    D_i(x) = I\Bigl\{{m^r(x,\eta_i) \ge U_i^r}\Bigr\}
\end{equation}
where  $m^r( x,\eta_i):= F_{\nu^r \vert \eta} \Bigl({S^r \left({x,\eta_i}\right)}\Bigr)$ and $U_i^r \vert \eta_i \sim \mathcal{U}[0,1]$. $m(x,\eta)$ is the take-up rate of or the objective demand function for option 1 for individuals with unobserved attributes $\eta$, when observed attributes are exogenously set to $x$. Similarly, define:
\begin{equation}\label{eq:definition_P}
m(x,\eta_i):= F_{\nu \vert \eta} \Bigl({S \left({x,\eta_i}\right)}\Bigr)
\end{equation}
the take-up rate or the stated demand function in the stated choice analysis.

Both Equations (\ref{eq:definition_D}) and (\ref{eq:definition_P}) share one important feature: they both depend on $\eta_i$. Crucially, the unobserved heterogeneity affecting the actual choice needs to be a sub-vector of the unobserved heterogeneity influencing stated choices. To simplify the exposition, the same vector is used in both cases. The presence of $\eta_i$ in both stated and actual demand functions is, I propose, the key link between stated and actual choices, and one of the main goals in eliciting choice probabilities in hypothetical scenarios. In fact, the identification results proposed in the next section would remain valid with any survey experiment that would guarantee that (1) individual choices within the experiment are affected by the same unobserved heterogeneity vector $\eta_i$ as in the actual choice, and (2) this individual heterogeneity can be recovered from the survey experiment. The case for using stated preferences over the hypothetical scenarios mimicking the actual choice environment is both intuitive and compelling. Even if they only replicate actual choices imperfectly, carefully designed stated preference analyses induce respondents to consider the variables that influence their decision in a real environment. Ideally, these variables are also reflected in their stated choice. The next section shows how to recover the unobserved heterogeneity from elicited choice probabilities.

In both this and the next section, the issue of measurement error is intentionally omitted for simplicity; however, Section \ref{sec:measurement_error} discusses identification with measurement error. Section \ref{subsec:main_identification} below discusses the main identification result. 

\subsection{Main Identification Result}\label{subsec:main_identification}
The difference $m^r(X_i,\eta_i) - m(X_i,\eta_i)$ measures the bias between the actual and the stated choice for individuals with choice characteristic $(X_i,\eta_i)$. More generally, one can define a mean-squared bias (MSB) as:
\begin{equation}\label{eq:definition_MSB}
    MSB(x) := \mathbb{E} \Bigl[{\bigl({m^r(X_i,\eta_i) - m(X_i,\eta_i)}\bigr)^2 \Bigl|{ X_i = x}\Bigl.}\Bigr]
\end{equation}
MSB provides a more thorough assessment of the accuracy of stated choice than comparing subgroups with observed characteristics $X_i=x$ via the difference $\mathbb{E}(D_i\vert X_i=x) - \mathbb{E}(P_{it} \vert X_i=x)$. 

The analyst may also be interested in a counterfactual analysis in the spirit of \cite{chernozhukov2013}. For example, assume there are two groups in the population $g_1$ and $g_2$. $m_g^r(\cdot,\cdot)$ is the demand function in group $g \in \{g_1,g_2\}$ and $F_{{X,\eta}\vert g}$ is the distribution of $(X_i,\eta_i)$ in group $g$. The analyst is interested in the counterfactual distribution:
\begin{equation}\label{eq:definition_CD}
    F_{D\langle {1,2}\rangle} := \int_{\mathcal{X} \times \mathcal{H}} m_{g_1}^r(x,n) dF_{{X,\eta}\vert g_2} (x,n) 
\end{equation}
which gives the demand that would prevail if  members of group $g_1$ had the same distribution of characteristics, $(X_i,\eta_i)$, as members of $g_2$.

Finally, the analyst may be interested in features of the structural function $x \mapsto m^r(x,.)$, the objective demand function, such as the average structural function:
\begin{equation}\label{eq:definition_ASF}
    \mu^r(x) := \int_\mathcal{H} m^r(x,n) dF_{\eta}(n)
\end{equation}
The key identifying assumption for this counterfactual quantity is the following:
\begin{assumption}\label{ass:independence}
    Conditional on $\eta_i$, the resolved uncertainty, $\nu_i^r$, is independent of $X_i$, that is, $$\nu_i^r \independent X_i \vert \eta_i.$$
\end{assumption}
Assumption \ref{ass:independence} imposes that the unobserved heterogeneity $\eta_i$ is the source of endogeneity of $X_i$ in a regression of $D_i$ on $X_i$. Controlling for $\eta_i$ would purge the estimated demand function from the omitted variable bias. Thus, Assumption \ref{ass:independence} requires that, when reporting their intended choice, respondents account for factors that would create a statistical dependence between observed choice attributes and the resolved uncertainty. For example, if both future unobserved shocks on the utility of a STEM degree and STEM degree attributes depend on individual's ability in math-related subjects, a respondent should factor-in their innate ability in reporting their intended choice.

In the rest of the paper, the analyst is assumed to elicit stated preferences in $T+1$ hypothetical scenarios. Hypothetical scenarios refer to exogenously set values of the variable $X$. The first scenario denoted $0$ serves for identification of the stated demand function. In this scenario, $X_{i0}$ spans the support of $X_i$. The stated demand function is needed for identification of $MSB(x)$ only. The remaining $T$ scenarios serve to learn about the unobserved heterogeneity $\eta$. The reader should think of them as either drawing random values of $X_i$, or setting a common counterfactual for all individuals.

To understand the intuition of identification, it is easier to proceed first with the case where $d:= \text{dim}(\eta) =1$. Assume further that the function $\eta \mapsto m(x,\eta)$ is strictly monotone. Given the flexibility for the researcher to analyse counterfactual scenarios, assume that $X_{i1}=\bar{x}$ for any individual $i$, that is, one counterfactual elicits the intended choice in a common hypothetical scenario. It entails that $P_{i1} = m(\bar{x},\eta_i)$. Because the counterfactual $\bar{x}$ is the same for all individuals, there is a one-to-one mapping between the stated choice and the unobserved heterogeneity. Observing $P_{i1}$ is equivalent to observing $\eta_i$. Hence:
\begin{equation}\label{eq:proof}
\begin{array}{rll}
    \mu^r(x) &=_{(1)}& \int_\mathcal{H} m(x,n ) dF_{\eta}(n)\\
    &=_{(2)}& \int_\mathcal{H} \int_0^1 I\bigl\{{m(x,n ) > u}\bigr\} dF_{U^r \vert \eta_i}(u\vert n)\; dF_{\eta}(n)\\
    &=_{(3)}& \int_\mathcal{H} \int_0^1 I\bigl\{{m(x,n ) > u}\bigr\} dF_{U^r \vert X_i,\eta_i}(u\vert x,n)\; dF_{\eta}(n)\\
    &=_{(4)}& \int_\mathcal{H} \mathbb{E}(D_i  \vert X_i=x,\eta_i=n ) dF_{\eta}(n)\\
    &=_{(5)}& \int_{[0,1]} \mathbb{E}(D_i  \vert X_i=x,X_{i1}=\bar{x}, P_{i1} = p ) dF_{X_{i1},P_{1}}(\bar{x},p)
\end{array}
\end{equation}
where $\mathcal{H}$ is the support of $\eta$. The first equality is by definition, the second equality follows from the fact that the conditional distribution of $U^r$ is uniform. Equality (3) uses Assumption \ref{ass:independence}. Equality (4) is by definition. Equality (5) follows from the fact that there is a one-to-one mapping between $P_{i1}$ and $\eta_i$ for all those such that $X_{i1}=\bar{x}$. 

A similar result holds for the MSB:
\begin{equation}\label{eq:proof_MSB}
\begin{array}{ll}
    MSB(x)&= \int_{[0,1]} \Bigl({\mathbb{E}(D_i  \vert X_i=x,X_{i1}=\bar{x}, P_{i1} = p ) }\Bigr.\\
    & \hskip64pt \Bigl.{- \mathbb{E}(P_{i0}  \vert X_{i0}=x,X_{i1}=\bar{x},P_{i1} = p ) }\Bigr)^2  dF_{X_{i1},P_{1} \vert X_i}( \bar{x},p \vert x).
    \end{array}
\end{equation} 
and $F_{D\langle {j,k}\rangle}$:
\begin{equation}\label{eq:proof_Fjk}
F_{D\langle {j,k}\rangle} = \int_{[0,1]} \mathbb{E}(D_i  \vert X_i=x,X_{i1}=\bar{x}, P_{i1} = p, i \in g_j ) dF_{X_{i1},P_{1} \vert X_i}( \bar{x},p \vert x, i \in g_k).
\end{equation} 
In sum, the fundamental idea is to match individuals on their stated preferences. The logic is the same for an heterogeneity in higher dimensions. 
\begin{assumption}\label{ass:invertibility}
Let $\dim(\eta)=d$.\\ (1) There are $T+1 > d$ scenarios, such that:
\begin{equation}\label{eq:system}
    P_{it} = m(X_{it},\eta_i), t=0,1,\ldots,T\\
    \text{ with } \{X_{it}: t=1,\ldots,T\} \independent \eta,
\end{equation}
In particular, we may have  $X_{it} = \bar{x}_t$ for all $i$. The support of $X_{i0}$ is $\mathcal{X}$. The support of $X_{it}, t \ge 1$ is $\mathcal{C}$, is a compact included in the support of $X_i$, $\mathcal{X}$.\\ (2) Furthermore, for any $\boldsymbol{X} \in \mathcal{C}^d$, such that $\textrm{rank}(X)=d$, the mapping $\eta \mapsto m(\boldsymbol{X},\eta) \in [0,1]^d$ is \textit{globally homeomorphic}, that is continuous and one-to-one.
\end{assumption}
Assumption \ref{ass:invertibility}.(1) imposes that the analyst observes at least as many scenarios as the dimension of $\eta$. In hypothetical scenarios, observed choice attributes are either (i) randomly generated from a set support, or (ii) fixed for everyone in the population. Note that the main objective of these hypothetical scenarios is to learn about the distribution of unobserved heterogeneity and not necessarily about the stated demand function.

Assumption \ref{ass:invertibility}.(2) is key. It imposes that given $\boldsymbol{X}\in \mathcal{C}^d$, such that $\textrm{rank}(X)=T=d$, a unique $\eta$ can be recovered from $\boldsymbol{P} = m(\boldsymbol{X},\eta).$ 
One example often used in empirical applications is a multiplicatively separable log-odd model, of which the model of \cite{blass2010} is a special case. It corresponds to:
\begin{equation}
\boldsymbol{P} = \Gamma \left({v(\boldsymbol{X} + K(\boldsymbol{X}) \eta )}\right)
\end{equation}
where $\Gamma(x) = \exp(x)/(1 + \exp(x))$, and $K(\boldsymbol{X})$ is $d \times d$ matrix with full rank that spans $\mathcal {X}^d$. In this case, $\eta = K(X)^{-1} \left({\Gamma^{-1}(\boldsymbol{P}) - v(\boldsymbol{X})}\right)$.

Global invertibility conditions is used, for example, in \cite{matzkin2008} to show global identification for general nonparametric simultaneous equation models. A detailed derivation of sufficient conditions on individual preferences is beyond the scope of this paper. However, the next paragraphs offer some comments. 

In the context of consumer choice, sufficient conditions on preferences for invertibility between demand and nonseparable individual heterogeneity are discussed, for example, in \cite{beckert2008}. Their framework is akin to that of this paper, and the derived conditions rest on the theorems of \cite{gale1965} and \cite{mas1979} for the existence of (global) homeomorphisms.\footnote{\cite{beckert2008} consider the demand, $\boldsymbol{x}$ for $d$ goods, characterised by a vector of prices $\boldsymbol{p}$, with $y$ being the agent income, and $\eta$ being the unobserved heterogeneity. $\boldsymbol{p}$ and $y$ are independent of $\eta$. The reduced form demand is characterised by a system $\boldsymbol{x} = m(\boldsymbol{p},y,\eta)$.} Following their discussion, the homeomorphism property must be deduced from properties of the Jacobian, $\nabla_{\eta} m(\boldsymbol{X},\eta)$.

Take for example an agent who perceives a utility $U_j(x_j,\eta) + \nu_j$ in option $j \in \{0,1\}$, for a resolved uncertainty $\nu_j$, with $U_j$ having the usual continuity and strict concavity properties. The utility gain from option 1 is then: $U_1(x_1,\eta) - U_0(x_0,\eta) + \nu_1 - \nu_0$. The stated preferences over $d$ hypothetical scenarios can therefore be summarised by the system:
\begin{equation}
    \boldsymbol{P} = m(\boldsymbol{X},\eta) = F_{\nu \vert \eta} \left({S(\boldsymbol{X},\eta) \vert \eta }\right)
\end{equation}
where $\nu:= \nu_0-\nu_1$, $F_{\nu \vert \eta} $ is the conditional distribution of the resolvable uncertainty as perceived by the agent, $S(\boldsymbol{X},\eta)=U_1(\boldsymbol{X}_1,\eta) - U_0(\boldsymbol{X}_0,\eta)$, and $\boldsymbol{X}= (\boldsymbol{X}_0, \boldsymbol{X}_1)$. The Jacobian matrix can be written:
\begin{equation}\label{eq:jacobian}
\begin{array}{cl}
      \nabla_{\eta} m(\boldsymbol{X},\eta) =  & f_{\nu \vert \eta} \left({S(\boldsymbol{X},\eta) \vert \eta }\right) \bigl({\nabla_\eta S(\boldsymbol{X},\eta)}\bigr) + (\nabla_\eta F_{\nu \vert \eta}) \bigl({S(\boldsymbol{X},\eta) \vert \eta }\bigr)
\end{array}
\end{equation}
where $f_{\nu \vert \eta} $ is the probability distribution function of the resolvable uncertainty. Assuming that $m$ is continuous in $\eta$, a Jacobian matrix with full rank is sufficient for a local homeomorphism, that is local invertibility. A local homeomorphism is a sufficient condition in the case where scenarios are set at fixed value for the whole population. In addition, if $\eta$ is an interior solution and the Jacobian has a positive determinant, then $m$ is a global homeomorphism \citep{mas1979}.

When $\eta$ is unidimensional and the resolvable uncertainty does not depend on $\eta$, this requires $\eta \mapsto S(\boldsymbol{X},\eta)$ to be strictly monotone for any $X$. In other words, $\eta$ should have a monotonic effect on the differential of utility between option 0 and 1. Because the resolvable uncertainty may also depend on $\eta$, the monotonicity of $S$ in $\eta$ needs to be strong enough to compensate for any converse effect of $\eta$ on the distribution of the resolvable uncertainty. The global homeomorphism property generalises the monotonicity requirements.

The next proposition is the main result of paper.
\begin{proposition}\label{prop:identification}
    Suppose Assumption \ref{ass:invertibility} holds, and $T=d$. Then MSB(.) and $F_{D\langle {\cdot,\cdot}\rangle}$ are identified from the joint distribution of $\bigl({D_i,X_i,X_{i0}, X_{i1}, \ldots, X_{iT}, P_{i0}, P_{i1}, \ldots, P_{iT}}\bigr)$. If Assumption \ref{ass:independence} also holds, then $\mu^r(.)$ is identified from the joint distribution\\$\bigl({D_i,X_i, X_{i1}, \ldots, X_{iT},  P_{i1}, \ldots, P_{iT}}\bigr)$. For example:
    \begin{equation}
        \mu^r(x) = \int_{\mathcal{C}^T \times [0,1]^T} \mathbb{E}(D_i  \vert X_i=x, \boldsymbol{X}_{i}^T = \boldsymbol{x}^T,\boldsymbol{P}_{i}^T = \boldsymbol{p}^T ) dF_{\boldsymbol{X}^T,\boldsymbol{P}^T}(\boldsymbol{x}^T,\boldsymbol{p}^T).
    \end{equation}
where $\boldsymbol{X}_{i}^T = \left({X_{i1}, \ldots, X_{iT}}\right)$ and  $\boldsymbol{P}_{i}^T = \left({P_{i1}, \ldots, P_{iT}}\right)$.    
\end{proposition}
The proof follows the same steps as Equation (\ref{eq:proof}), \textit{mutatis mutandis}. Proposition \ref{prop:identification} summarises the main insight of the paper. By matching individuals on their stated preferences, the analyst can control for their unobserved heterogeneity. This allows assessing average forecast error by using a finer definition of heterogeneity or by performing counterfactual analyses. In addition, if conditioning on the unobserved heterogeneity ensures that the resolved uncertainty and the observed choice attribute are independent, the analyst can learn about causal effects of choice attributes.

\section{Identification in the presence of measurement error}\label{sec:measurement_error}
The previous section ignored measurement error to simplify the exposition. It is possible to allow for some form of classical measurement error while preserving the main identification result. This section deals with measurement errors that arise from respondents reporting inaccurately their true intended choice, possibly due to inattention, misunderstanding the survey instrument, or lack of effort. Thus, the measurement error is viewed as a source of randomness across scenarios. The case where measurement error would be systematic over all scenarios amounts to consider an additional dimension of unobserved heterogeneity, say $\xi$. For the purpose of causal inference, Assumption \ref{ass:independence} should be adapted to state that $\nu_i^r \independent X_i, \xi_i \vert \eta_i$.

Assumption \ref{ass:pseudo-panel-update} summarises the structure of the data available to the researcher.
\begin{assumption}[Pseudo-panel]\label{ass:pseudo-panel-update}
Let $t = 1,\ldots, T,$ be the index for $T$ hypothetical scenarios. The researcher observes an i.i.d. sample, $\{P_{it}^*, X_{it}\}_{i = 1,\ldots,N;t=0,1,\ldots,T}$ such that:
\begin{equation}\label{eq:perturbed_model}
\begin{array}{l}
L(P_{it}^*) = L(P_{it}) + \epsilon_{it}:= h(X_{it},\eta_{i}) + \epsilon_{it}, t = 0, 1, \ldots, T \end{array}
\end{equation}
with $\{X_{it}\}_{t=0,1,\ldots,T} \independent \eta_i$. In particular, we may have $X_{it} = \bar{x}_t$ for all $i$. The support of $X_{i0}$ is $\mathcal{X}$. The support of $X_{it}$, $\mathcal{C}$ is a compact included in the support of $X_i$, $\mathcal{X}$, and $L(\cdot)$ a known function, e.g. the logit function. $\eta_{i}$ is an unobserved heterogeneity of finite dimension $d \le T$, and $\epsilon_{it}$ is a continuously distributed random variable. The distribution function of $\epsilon_{it}$ obeys:
\begin{enumerate}
\item[(i)] $
f_{\epsilon_{t} \vert X_{t}, \eta, X_{(-t)}, \epsilon_{(-t)} } (u \vert x_{t}, \tau, x_{(-t)}, \epsilon_{(-t)}) = f_{\epsilon \vert X_{t}} (u\vert x_{t})$,\\ for all $t=0,1,\ldots, T$, and $(x_{t}, \tau, x_{(-t)}, \epsilon_{(-t)}) \in \mathcal{X}\times \mathbb{R} \times  \mathcal{X}^{T-1}\times \mathbb{R}^{T-1}$,
\item[(ii)] $\mathbb{E}(\epsilon_{it} \vert X_{it})= 0$ for all $i$ and $t=1,\ldots, T$,
\item[(iii)] $\phi_{\epsilon_{t} \vert X_{t}} (s \vert x)$ does not vanish for all $i$ and $t$, where $\phi_{Y \vert X}(s|x)$ is the (conditional) characteristic function of $Y$, for all $s \in \mathbb{R}$ and $x \in \mathcal{X}$.
\end{enumerate}
\end{assumption}
$P_{it}^*$ deviates from the true stated choice because of the measurement error $\epsilon_{it}$. Assumption \ref{ass:pseudo-panel-update} imposes that $\epsilon_{it}$ is a classical measurement error, that is, independent across scenarios, unrelated to $\eta_i$, but possibly related to $X_{it}$. The main identification idea is to use \cite{kotlarski1967}'s Lemma to show that the joint distribution of $(D_i,X_i,X_{i1},\ldots, X_{it}, P_{i1}, \ldots, P_{it})$,is identified from the joint distribution of $(D_i,X_i,X_{i0},X_{i1},\ldots, X_{it}, P_{i0}^*, P_{i1}^*,\ldots,P_{it}^*)$. The argument is reminiscent of \cite{evdokimov2010} for the case $d=1$ and \cite{meango2023}, for the case where $d>1$. 

Suppose that $d=1$, $t \in \{0,1\}$. The main goal is to show that the conditional distribution of $P_{i1}$ given $(D_i, X_i, X_{i1})$ is identified from the data. Indeed, since the joint distribution of $(D_i, X_i, X_{i1})$ is identified from the data, by Bayes rule, one can recover the quantity $\mathbb{E}(D_i\vert X_{i}, X_{i1}, P_{i1})$  that is needed for identification (see Equation (\ref{prop:identification})).  
Note that for any $i$ and an arbitrary $x$ such that $X_{i0} = X_{i1} = \bar{x}$:
\begin{align}
\begin{pmatrix}
L(P_{i0}^*)\\
L(P_{i1}^*)
\end{pmatrix}
& \bigg\rvert \left\{{X_{i0} = X_{i1} = \bar{x}}\right\} = \begin{pmatrix}
m(\bar{x},\eta_i) + \epsilon_{i0}\\
m(\bar{x},\eta_i) + \epsilon_{i1}
\end{pmatrix}
\end{align}
Lemma 1 of \cite{evdokimov2010} implies that the conditional characteristic distribution of $\epsilon_{it}$, say $\phi_{\epsilon_{t} \vert X_{it}}(s \vert \bar{x})$ is identified from the joint distribution of $(X_{i0},P_{i0}^*,X_{i1},P_{i1}^*)$ on the set such that $\{x \in \mathcal{X}: X_{i0}=X_{i1}=x\}$. Assume for simplicity that this set covers $\mathcal{C}$.\footnote{The flexibility of belief elicitation means that the researcher can purposefully induce a repetition of the same scenario at any point of the support.} Given identification of the characteristic function of measurement error, we can identify the characteristic functions:
\begin{equation}\label{eq:deconvolution_eta1}
\phi_{L(P_1)\vert D_i, X_i, X_{i1}}(s \vert d,x,x_1) =  \phi_{L(P_1^*)\vert D_i, X_i, X_{i1}}(s \vert d,x,x_1) /\phi_{\epsilon_{1}\vert X_{i1}}(s \vert x_1).   
\end{equation}
Knowledge of the characteristic function being equivalent to knowledge of the distribution function, this shows the result for $d=1$. For the case $d=2$, $T=2$, one can repeat the same steps and note that:
\[\begin{array}{lll}
    &&\phi_{L(P_1^*),L(P_2^*) \vert {D, X, X_{1},X_{2}}}(s_1,s_2\vert D_i, X_i, X_{i1},X_{i2}) \\
    &&= \mathbb{E} \left({\exp \left({i(s_1 L(P_{i1}^*) + s_2 L(P_{i2}^*) \vert D_i, X_i, X_{i1},X_{i2})}\right)}\right) \\
    &&=  \mathbb{E} \left({\exp \left({i(s_1 L(P_{i1}) + s_2 L(P_{i2}) \vert D_i, X_i, X_{i1},X_{i2} }\right)}\right)\cdot \mathbb{E} \left({\exp \left({i(s_1  \epsilon_{i1} + s_2 \epsilon_{i2} \vert D_i, X_i, X_{i1},X_{i2})}\right)}\right)\\
    &&= \phi_{L(P_1),L(P_2) \vert {D, X, X_{1},X_{2}}}(s_1,s_2\vert D_i, X_i, X_{i1},X_{i2}) \cdot \phi_{\epsilon_1\vert X_{i1}}(s_1 \vert X_{1})\phi_{\epsilon_2 \vert X_{2}}(s_2 \vert X_{i2})
\end{array}
\]
The same development can be repeated for $d>2$ \textit{mutatis mutandis}. Proposition \ref{prop:identification_with_error} summarises the result and updates Proposition \ref{prop:identification} for the case of measurement error.
\begin{proposition}\label{prop:identification_with_error}
    Suppose Assumption \ref{ass:invertibility} and \ref{ass:pseudo-panel-update} hold. $T\ge d$ and $\{x \in \mathcal{X}: X_{ij}=X_{ik}=x, \text{ for some } 0 \le j,k \le T\}=\mathcal{X}$. Then MSB(x), $x \in \mathcal{C}$, and $F_{D\langle {\cdot,\cdot}\rangle}$ are identified from the (observed) joint distribution of $\bigl({D_i,X_i,X_{i0}, X_{i1}, \ldots, X_{iT}, P_{i0}^*,P_{i1}^*, \ldots, P_{iT}^*}\bigr)$. If in addition, Assumption \ref{ass:independence} holds, $\mu^r(.)$ is identified from the same joint distribution.
\end{proposition}

Proposition \ref{prop:identification_with_error} shows that even a small number of scenarios allows identifying the joint distribution of unobserved heterogeneity. $MSB(x)$ is identified on $x \in \mathcal{C}$, because $P_{i0}^*$ is observed with error. The error can be corrected only on the set such that: $\{x \in \mathcal{X}: X_{i0}=X_{it}=x\} = \mathcal{C}$. Thus, the joint distribution of $P_0,X_i$ is only identified on $\mathcal{C}$. Identification on the full support requires $\mathcal{C}= \mathcal{X}$. 

The identification result being constructive, it can serve for a three-step estimation as in \cite{evdokimov2010} or \cite{meango2023}: first, a deconvolution estimator for the joint distribution of $\{P_{t}: t = 1, \ldots,T\}$ from the joint distribution of $\{P_{t}^*: t = 0,1, \ldots,T\}$. Second, an estimation of the joint distribution of $(D_i,X_i, P_1, \ldots, P_T)$. Finally, estimation of the parameters of interest by integration. However, these steps can be involved and practitioners are unfamiliar with such tools. The next section proposes instead to assume that $T$ is large relative to $d$ and use the Two-Step Group Fixed Effects estimator proposed by \cite{bonhomme2022}.

\section{Estimation}\label{sec:estimation}
This section proposes to use the Two-step Group Fixed-Effect (TSGFE) methodology of \cite{bonhomme2022} for estimating $MSB(\cdot)$, $F_{D\langle{\cdot,\cdot}\rangle}$ and $\mu_r(\cdot)$, with data that satisfy Assumptions \ref{ass:invertibility} and \ref{ass:pseudo-panel-update}. Under the latter, their crucial Assumption 2 of existence of injective moments is satisfied, for $m(\cdot,\cdot)$ being Lipschitz-continuous and the variance of $\epsilon_{it}$ being finite. The procedure consists in two steps: the first step involves a classification that uses auxiliary moments to classify individuals into types, based on a finite (low) dimensional unobserved heterogeneity. The second step uses the estimated classification to estimate the parameters of interest. In the setting of this paper, TSGFE  is a natural approach to exploit the link between probabilistic stated choices and actual choices through the unobserved heterogeneity $\eta$. This section assumes that $T>\dim(\mathcal{X})$ is large.

\paragraph{First-step: Classification by kmeans clustering.} The procedure relies on individual-specific moments $h_i$. Identification in our framework requires at least $d$ moments. It is as follows:\\
Estimate consistently:
\begin{equation}
    \widehat{\mathbb{E}}\bigl({L\left({P_{it}^* \vert X_{it}}\right)}\Bigr) = h_{i}(X_{it}),
\end{equation}
for each individual $i$; generate (at least) $d$ moments:
\begin{equation}
    h_{ij} = h_{i}(\bar{x}_j), j \in \left\{{1,\ldots,d}\right\},
\end{equation}
where the rank of the matrix that stacks the vectors $\bar{x}_j$ is (at least) $d$; finally, partition individuals into $K$ groups, corresponding to group indicators $\widehat{k}_i \in \left\{{1,\ldots,K}\right\}$ by computing
\begin{equation}\label{eq:kmean_problem}
    \left({\widehat{h}(1), \ldots, \widehat{h}(K), \widehat{k}_1, \ldots, \widehat{k}_N}\right) = \argmin\limits_{\left({\tilde{h}(1), \ldots, \tilde{h}(K), k_1, \ldots, k_N}\right)} \sum_{i=1}^N \left\|{h_i - \tilde{h}(k_i)}\right\|^2,
\end{equation}
where $\{k_i\}$ are partitions of $\{1,\ldots,N\}$ into $K$ groups, and $\tilde{h}(K)$ is a vector of dimension $d$. \cite{bonhomme2022} provide a data-driven selection rule for the choice of $K$.

\paragraph{Second-step: (Parametric) estimation.} The TSGFE uses a parametric form for Equation (\ref{eq:definition_D}). Assume that there exists a finite dimension parameter $\theta_0$, such that $m(X_i,\eta_i) = m(X_i,\eta_i; \theta_0)$. In the second step, the estimator maximises the likelihood function $\ln \Pr(D_i \vert X_{i}, \eta_i;\theta_0)$ with respect to the common parameter $\theta$ and group-specific effects, that is:
\begin{equation}
    \left({\widehat{\theta},\widehat{\eta}(1),\ldots, \widehat{\eta}(K)}\right) = \argmax\limits_{\left({\theta,\eta(1),\ldots, \eta(K)}\right) } \sum_{i=1}^N \ln \Pr(D_i \vert X_{i}, \eta(\widehat{k}_i);\theta)
\end{equation}
Estimates of $\mu^r(\cdot)$, $F_{D\langle{\cdot,\cdot}\rangle}$ and $MSB(\cdot)$ follow naturally by taking empirical means. For example, if $\widehat{p}_k$ is the proportion of individuals in group $k$, the estimator for $\mu_r(x)$ is given by:
\begin{equation}
    \widehat{\mu^r}(x) = \sum_{k=1}^K \widehat{p}_k \sum_{i:\widehat{k}_i \in k} m\left({X_i,\widehat{\eta}(\widehat{k}_i);\widehat{\theta}}\right)
\end{equation}
\cite{bonhomme2022} discuss regularity conditions and asymptotic properties of TSGFE in appropriate length. Appendix \ref{app:simulations} provides simulation results that demonstrates the usefulness of the method in our context.\footnote{A routine easily adaptable is also available at the following \href{https://www.dropbox.com/scl/fi/ri53cxa76lt5bkgd7qowy/simulation_RM.m?rlkey=voz2bz979rq1tgjh21pv1zbeg&dl=0}{link} }

\section{Discussion}\label{sec:discussion}
The proposed strategy is of practical importance to empirical researchers for two types of analyses: (1) causal inference, and (2) heterogeneity analyses.

Causal inference often requires exogenous variations that may not be available in some contexts. Even when a suitable instrument is available, it recovers treatment effects for sub-populations that may differ from that of the entire population. One example is a situation where the researcher would only have access to administrative data, which records individual choices and a limited set of observed characteristics. The above strategy suggests complementing these data with a survey experiment, where choice probabilities are elicited and the main treatment of interest is varied along with other relevant characteristics. Provided that the unobserved heterogeneity recovered is the main source of bias, the analyst can estimate (marginal) treatment effects for the entire population. The latter hypothesis is of course untestable, as are the critical assumptions of the other existing strategies. Still, the researcher can test whether the survey experiment captures all the information on the unobserved heterogeneity available from the stated choice (See Appendix \ref{app:testing}). If a surrogate for an experiment exists (exogenous variation or discontinuity at a threshold), the researcher can validate the obtained results using the stated preference approach. It suffices to compare them with the results from using the surrogate experiment, on the sub-population where the latter permits identification of a causal effect.

The discussion above implies that the proposed strategy should not necessarily be viewed as a substitute for existing strategies, but as a complement. The ability to recover unobserved heterogeneity is valuable to (i) increase the credibility of an instrument, (ii) control for confounding factors that are not necessarily separable from treatment status, or (iii) assess balance across a threshold. In the case of randomised control trials, identifying the unobserved heterogeneity may help to understand treatment effect heterogeneity, or improve the precision of the estimator. Hence, even in situations where usual inference strategies are available, a stated preference analysis may aid inference. Thus, stated preference can complement and enhance the applied researcher toolbox.

To the best of my knowledge, two recent contributions exploit the idea that a stated preference analysis can help in recovering some information on individual heterogeneity and performing causal inference: \cite{briggs2020} and \cite{bernheim2022}. \cite{briggs2020} adapt the generalized Roy model to show that marginal treatment effects are identified when data on stated preferences are available. Translated into our framework, $D$ would be a sector choice (the treatment), and the effect of interest would pertain to a third variable $Y$. Their framework imposes additive separability of $\eta$ and $X$, and, in allowing  $\eta$ to be considered as univariate, they correct for the selection bias from the observation of one single intended choice. However, point identification breaks down in the case of measurement error. This paper complements their work and shows that, with repeated observations, matching on a range of stated preferences corrects for the selection bias, even in the presence of nonseparable heterogeneity and measurement error.

\cite{bernheim2022} characterises the relation between stated preferences and realised outcomes. Under the assumption that this relation is stable across treatment, the treatment effect revealed from the stated preference analysis can be used to infer the treatment effect on the actual choice. A key advantage of their methodology is that it does not require jointly recording choices and stated preferences in the same population. Furthermore, stated choices can be binary and $\eta$ can remain unrestricted. The key exclusion restriction is that treatment can only affect actual choice through the stated preference. In contrast, the assumptions in this paper on the relationship between stated and actual choice are significantly milder. The main requirement is for the unobserved heterogeneity in the actual choice to be a subset of the unobserved heterogeneity in the choice experiment. Thus, the strategy used here complements their approach in the case where the exclusion assumption fails.

\section{Conclusion}\label{sec:conclusion}
Stated choices can serve to understand actual behaviour, not by matching actual choices, but rather by providing valuable information on individual heterogeneity. This paper shows how to harness stated preference analyses to recover the distribution of individual unobserved heterogeneity (i.e. agent's types). Once recovered, the types serve to evaluate biases in prediction for subgroups. If they are the main source of endogeneity, their introduction corrects for biases in demand function estimations.

Up to now, stated and revealed preference analyses have mostly been conducted separately due to the strong assumptions needed to combine them. The hope is that this research will convince practitioners that stated preference analyses can become part of their toolbox for valid counterfactual analyses of actual choices. 

\bibliographystyle{apalike}
\bibliography{ref_expectations}
\newpage
\section*{Appendix}
\appendix
\section{Testing for the Dimension of $\eta$}\label{app:testing}
This section addresses the question of the dimensionality of $\eta$. A restriction on the dimension of unobserved heterogeneity has testable implications. Starting from the case of an univariate heterogeneity is useful. A one-dimensional heterogeneity imposes that a person who is relatively prone to choose one option in a given scenario will always be relatively prone to chose the same option in all other hypothetical scenarios. This property, known as rank invariance, has been introduced by \cite{doksum1974}, and used more recently by \cite{heckman1997}, \cite{chernozhukov2005}, and \cite{torgovitsky2015}, among others. Violation of rank invariance suggests an heterogeneity with dimension of at last two. This logic can be repeated: if after matching on $d$ probabilistic choices, rank invariance is still violated, then the dimension of unobservable heterogeneity is at least $d+1$.

Taking the case $d=1$, $T=2$ it is easy to see that, without measurement error, for any $x_1,x_2 \in \mathcal{X}$, the rank of an individual $i$ is invariant, irrespective of the value of $X_{it}$, and is given by $F_{\eta}(\eta_i)$. Thus, the hypothesis:
\[ \text{H}_0: F_{P_1 \vert X_{1}  }(P_{i1} \vert X_{i1}) = F_{P_2\vert X_{2} }(P_{i2} \vert  X_{i2})\]
provides a testable implication of rank invariance. If it is rejected, an additional degree of freedom is required to rationalised the data.

When accounting for the possibility of measurement error, the equality under $\text{H}_0$ breaks down because rank slippage may appear due to the measurement error. Yet, measurement error does not impair our ability to estimate the conditional joint distribution of $(P_{i1},P_{i2})$, as demonstrated in Section \ref{sec:measurement_error}. If $\eta$ is unidimensional, the conditional distribution of $P_2$ satisfies:
\begin{equation}\label{eq:distribution_P2}
    F_{P_2 \vert X_{2}, X_{1},P_{1}} (p_2 \vert X_{i2},X_{i1},p_1)= \left\{{\begin{array}{cl}
         0 & \text{ if } 0 \le p_2 < m(X_{i2},\eta_1(X_{i1})) \\
         1 & \text{ otherwise.}
    \end{array}
    }\right.
\end{equation}
where $\eta_1(X_{i1})$ is such that: $p_1 = m(X_{i1},\eta_1(X_{i1}))$. Equation \ref{eq:distribution_P2} implies that conditioning on $(X_{i2}, X_{i1},P_{i1})$ determines the value of $P_{i2}$. A testable implication of the restriction $\dim(\eta)=1$ is therefore:
\[ \text{H}_0: Q_{P_2 \vert X_{2}, X_{1},P_{1}} (\overline{\tau} \vert X_{i2},X_{i1},p_1) = Q_{P_2 \vert X_{2}, X_{1},P_{1}} (\underline{\tau} \vert X_{i2},X_{i1},p_1), \text{ for any } 0< \underline{\tau}< \overline{\tau} < 1.\]
It is straightforward to repeat the same logic for higher dimensions.

\section{Simulation Details}\label{app:simulations}
The proposed simulation performs inference on the following simulated data:
$X_i$ represents the observed characteristics. $(\eta_1,\eta_2)$ is a two dimensional vector that influences both stated and actual choices. $\Phi(\nu_i^r)$ is the resolved uncertainty, where $\Phi$ is the normal density distribution. The DGP for $(X_i,\eta_1,\eta_2,\nu_i^r)$ is a joint normal distribution with mean $(1,0,0,0)$ and variance-covariance: \[
\Omega = \left[{\begin{array}{cccc}
     1 & 0.1 & 0.2 & 0 \\
       & 1  & 0.1 & 0.05 \\
       &   & 1 & 0.05 \\
       &   &   & 1
\end{array}}\right]\] 
The actual choice is generated through:
\[
D_i = I\Bigl\{{0 + 1 \;X_{i} + 0.5  \;\eta_{i1} + 0.1 \; X_{i}\; \eta_{i1} + \eta_{i2} > U_{i}}\Bigr\}
\]
Thus $d=2$. The stated preferences are generated using the log-odd function:
\[
\ln \left({ \dfrac{P_it}{1-P_it}}\right) = 0.1 + 0.8 \; X_{it} + 0.6 \; \eta_{i1} + 0.3 \; X_{it}\; \eta_{i1} + \eta_{i2} + \epsilon_{it}
\]
where $X_{it}$ are randomly drawn from a subset of $11$ points in the set $\{-2,-1.8, \ldots, 2\}$, and $\epsilon_{it} \sim \mathcal{N}(0,0.05 \exp(X_{it}))$. To mimic actual stated preference data, $P_{it}$ is rounded to the first decimal.
The simulation consider $S=1,000$ samples of size $N=1,000$, for $T=5,10,20$. In the first step, two moments are generated for $x=0$ and $x=1$. The parametric estimation in the second step performs a probit estimation, with group-specific intercepts and slopes.
\begin{figure}[htbp]
\centering
\caption{Simulation results for the treatment effect $\mu^r(1) - \mu^r(0)$}
\centering
\includegraphics[scale=1]{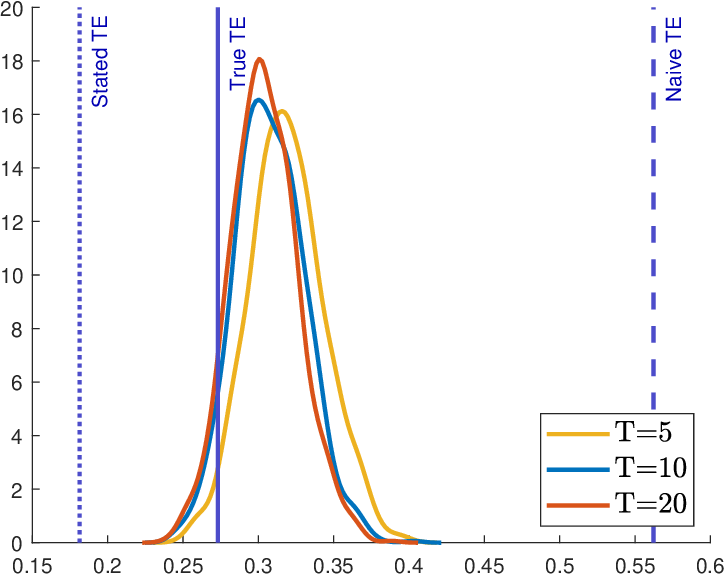} 
\label{fig:density}
\end{figure}

Figure \ref{fig:density} summarises the simulation results. It represents the density of the estimated treatment effect $\mu^r(1) - \mu^r(0)$ for different values of $T$. The true treatment effect 'true TE'=0.27 differs from the stated treatment effect 'Stated TE'=0.18 and the naive treatment effect 'Naive TE'=0.56. The estimated treatment effect using TSGFE, although sightly biased, correct appropriately for the omitted variable bias.  

\end{document}